\documentclass[aps,pra,preprint,nofootinbib,groupedaddress]{revtex4-1}
\usepackage{amssymb}
\usepackage{amsmath}
\usepackage{amsfonts}
\usepackage{color}
\usepackage{graphicx}
\usepackage{bm}
\usepackage{braket}
\usepackage{enumerate}

\usepackage[dvipdfmx,colorlinks=true]{hyperref}

\usepackage{amsthm}
\usepackage{here}

\newcommand{\sps}[1]{\ensuremath{^{\mathrm{#1}}}}

\newtheorem*{definition*}{Definition}

\begin{document}

\title{Quantum Information Capsule in Multiple-Qudit Systems and Continuous-Variable Systems}
\author{Koji Yamaguchi}
\author{Masahiro Hotta}
\date{\today}

\affiliation{Graduate School of Science, Tohoku University,\\ Sendai,
980-8578, Japan}

\begin{abstract}
Quantum correlations in an entangled many-body system are capable of storing information. Even when the information is injected by a local unitary operation to the system, the entanglement delocalizes it. In a recent study on multiple-qubit systems, it is shown that a virtual qubit defined in the correlation space plays a role of perfect storage of delocalized information, which is called a quantum information capsule (QIC). 
In order to enhance the capacity of quantum information storage, it is crucial to formulate the cases for multiple-qudit systems and continuous-variable (CV) systems. We analytically prove that it is possible to construct a QIC for general write operations of the systems. It turns out that the extension to quantum field theory is achievable.
For Gaussian states, we explicitly construct a QIC for shift write operations. 
We analyze the time-evolution of QIC in a CV system to demonstrate the diffusion of information in entangled pure states.
\end{abstract}

\maketitle

\section{Introduction}
Quantum information is a subtle concept. We have two reasons to consider it at present, even though this word has already been widely used. The first reason is related to the information loss problem in black hole evaporation. Black holes can emit the Hawking radiation \cite{HR}. During this process, the state might evolve from pure to mixed \cite{BHIP}. If so, unitarity of the quantum theory would be broken. Many researchers often say, ``Quantum information can be lost''. However, nobody has provided the precise definition of the quantum information in this context. We should sharpen the concept. The second reason of the subtlety is related to delocalization of information due to entanglement. For instance, consider a Bell state depending on a continuous real parameter $\theta$: $\ket{\text{Bell}(\theta)}\equiv \frac{1}{\sqrt{2}}\left(e^{-i\theta}\ket{0}\ket{0}+e^{i\theta}\ket{1}\ket{1}\right)$. By taking the partial trace, we get a reduced state for each qubit, which is independent of $\theta$. The information is delocalized in the system. This feature is crucial when we treat quantum devices of information storage. Where is the quantum information stored in the system? This is a very profound and important question. It should be handled with care. 

In order to address the above problems, one of our main tasks by using an entangled many-body system is storage and retrieval of information about continuous unknown parameters. In this paper, we define quantum information as classical information about parameters stored in a delocalized way. By assuming that the parameter $\theta$ is unknown, the above Bell state $\ket{\text{Bell}(\theta)}$ exhibits an example of quantum information since $\ket{\text{Bell}(\theta)}$ is also unknown and a superposition of two distinct states. 

There are two classes of ways to retrieve the quantum information. The common one is to use a pair of partners. A pair of subsystems in a many-body system is called partners when the pair is in a pure state. The notion of partners has originally been proposed in the analysis on modes of scalar field in the context of black hole physics \cite{HSU}. The partner of a specific mode is identified in \cite{HSU}, and a general formula for partners is obtained in \cite{TYH}. The definition of partners is extended to multiple-qubit systems in \cite{YWH} by using virtual qubits in the correlation space \cite{CS1,CS2}. It has been shown that for an arbitrary multiple-qubit pure state, we can identify the partner of the first qubit on which information is injected \cite{YWH}. Since the pair of partners is in a pure state both before and after the write operation, it perfectly confines the information of $\theta$. In order to retrieve the information out of the system by a SWAP operation, we need two external qubits. 
The second way to retrieve the delocalized information is to use quantum information capsule (QIC). A QIC is a single virtual qubit that perfectly confining the imprinted information in a pure state. It has been shown that for the multiple-qubit system in an arbitrary state, a QIC always exists \cite{YWH}. By swapping the state of a QIC for the state of a single external qubit, the information of $\theta$ is retrieved. After the SWAP operation, no information remains in the multiple-qubit system.

In \cite{YWH}, the QIC analysis was done only for multiple-qubit systems. However, we have many other physical systems in which we can implement storage and retrieval of information, such as multiple-qudit systems and continuous-variable (CV) systems. 
Information storage capacity can be enhanced by adopting a multiple-qudit system or a CV system.
In this paper, we analytically prove the existence of QIC's in the systems. QIC experiments can be expected by using entangled cold atoms in a pure state \cite{E1,E2,E3,E4}. 

Information stored in quantum systems is generally scrambled by the time-evolution. We demonstrate the diffusion of information keeping unitarity by numerically calculating the time-evolution of a QIC in a CV system. 

In Section \ref{sec_fin_dim}, we show that a QIC always exists for arbitrary states and write operation in multiple-qudit systems. A QIC stores $(d-1)$ independent parameters. In Section \ref{sec_inf_dim}, we extend the proof to CV systems. The results are applicable to scalar field, as long as the continuum limit can be taken properly. In Section \ref{sec_ex}, we demonstorate scrambling of information due to the free evolution of the system by calculating the time-evolution of a QIC. Summary is given in Section \ref{sec_summary}. 

In this paper, we adopt the natural units: $c=\hbar=1$.

\section{QIC in finite dimensional systems}\label{sec_fin_dim}
%Write operation 
In this section, we investigate delocalized information in an $N$-qudit system $\mathcal{H}=\mathcal{H}_d^{\otimes N}$, where $\mathcal{H}_d$ is a $d$-dimensional Hilbert space. 
Let us imprint the information of an unknown real parameter $\theta$ into the $N$-qudit system by performing a unitary write operation $\hat{W}(\theta)\equiv e^{-i\theta\hat{T}}$, where $\hat{T}$ is assumed to satisfy $\hat{T}=\hat{U}^\dag \left(\hat{t}\otimes \mathbb{I}^{\otimes N-1}\right)\hat{U}$ for a unitary operator $\hat{U}$ and $\hat{t}\in\mathfrak{su}(d)$ satisfying a normalization condition: $\mathrm{Tr}_{\mathcal{H}_d}\left(\hat{t}^2\right)=d$. If the system is initially in a pure state $\ket{\Psi}$, it evolves into $\hat{W}(\theta)\ket{\Psi}$.
When $\hat{U}=\mathbb{I}_d^{\otimes N}$, the write operation is a local unitary operation on the first qudit. The imprinted information is delocalized due to the entanglement in general. For $\hat{U}\neq \mathbb{I}_d^{\otimes N}$, the write operation is not necessarily a local operation. A nontrivial unitary operator $\hat{U}$ smears the write operation, as we will see in Section \ref{sec_inf_dim}.

%Fisher info for a parameter
After the write operation, the system evolves into $\ket{\Psi(\theta)}\equiv \hat{W}(\theta)\ket{\Psi}$. The quantum Fisher information
\begin{align}
 F\equiv 4\left(\Braket{\partial_\theta \Psi(\theta)|\partial_\theta \Psi(\theta)}-\left|\Braket{\Psi(\theta)|\partial_\theta \Psi(\theta)}\right|^2\right)=4\Braket{\Psi|\left(\Delta \hat{T}\right)^2|\Psi}
\end{align}
quantifies the best possible precision of the estimation of $\theta$ from the state $\ket{\Psi(\theta)}$ \cite{QF}. Here, we have defined $\Delta \hat{T}\equiv \hat{T}-\Braket{\Psi|\hat{T}|\Psi}$. It should be noted that $F$ is independent of $\theta$ in this setup. The information of $\theta$ is imprinted unless $F= 0$. 

%information delocalization
As a special example, let us consider the case when $\hat{U}=\mathbb{I}_d^{\otimes N}$. The write operation $\hat{W}(\theta)$ corresponds to a local unitary operation on the first qudit $\hat{w}(\theta)\otimes \mathbb{I}_d^{\otimes N-1}$, where we have defined $\hat{w}(\theta)\equiv e^{-i\theta\hat{t}}$.
Since the information of $\theta$ is imprinted by a local unitary operation of the first qudit, it can be extracted from the first qudit if the first qudit does not share entanglement with other qudits. Suppose that the initial state is given by $\ket{\Psi}=\ket{\phi}\otimes \ket{\psi}\in\mathcal{H}_d\otimes \mathcal{H}_d^{\otimes N-1}$, where $\ket{\phi}$ and $\ket{\psi}$ are unit vectors. After the write operation, it evolves into $\ket{\Psi(\theta)}=\ket{\phi(\theta)}\otimes \ket{\psi}$, where $\ket{\phi(\theta)}\equiv \hat{w}(\theta)\ket{\phi}$. The information of $\theta$ can be retrieved perfectly by swapping the state of the first qudit for the state of an external single qudit state. The SWAP unitary operation \cite{NC} is given by
\begin{align}
 \hat{U}_{\mathrm{swap}}=\sum_{i,j=1}^d\left(\ket{i}\bra{j}\otimes \mathbb{I}_d^{\otimes N-1}\right)\otimes  \ket{j}\bra{i}:\mathcal{H}\otimes \mathcal{H}_d^{\mathrm{(ext.)}}\to \mathcal{H}\otimes \mathcal{H}_d^{\mathrm{(ext.)}},
\end{align}
where $\left\{\ket{i}\right\}_{i=1}^d$ is an orthonormal basis for a single-qudit system and $\mathcal{H}_d^{\mathrm{(ext.)}}$ is $d$-dimensional Hilbert for the external qudit system. By using the set of generators $\left\{\hat{t}_{i}\right\}_{i=1}^{d^2-1}$, it can be rewritten as
\begin{align}
 \hat{U}_{\mathrm{swap}}=\frac{1}{d}\sum_{\mu=0}^{d^2-1}\left(\hat{t}_{\mu}\otimes \mathbb{I}_d^{\otimes N-1}\right)\otimes \hat{t}_{\mu}\label{eq_swapprod}
\end{align}
as is shown in Appendix \ref{app_swap}. Here, we have defined $\hat{t}_0\equiv\mathbb{I}_d$. Assuming the initial state of the external qudit system is given by $\ket{\chi}$, 
\begin{align}
 \hat{U}_{\mathrm{swap}}\ket{\Psi(\theta)}\otimes \ket{\chi}=\left(\ket{\chi}\otimes \ket{\psi}\right)\otimes \ket{\phi(\theta)}
\end{align}
holds. This means that the information of $\theta$ is confined in the first qudit. No information is left in the $N$-qudit system after the SWAP operation. However, when the first qudit is entangled with other qudits, the reduced state of the first qudit does not necessarily have perfect information of $\theta$ as we have seen in Introduction.

%Two pictures exist 
There are two known pictures of how the system stores the delocalized information in multiple-qubit systems with write operations $\hat{W}(\theta)=e^{-i\theta\hat{t}}\otimes \mathbb{I}_2^{\otimes N-1}$. 
A common one is that the first qubit and its purification partner shares it. The partner qubit is first defined in \cite{YWH}. The pair of entangled partners is in a pure state and stores the injected information of $\theta$. The second one is that a virtual qubit in the correlation space perfectly confines the delocalized information in a pure state \cite{YWH}. The virtual qubit is called a quantum information capsule (QIC). It should be noted that, in \cite{YWH}, these two pictures are investigated only in a special case of our setup: $\hat{U}=\mathbb{I}_d^{\otimes N}$ and $d=2$.

%Definition of virtual qudit
One of the main aims of this paper is to extend the notion of partners and QICs for general write operation  $\hat{W}(\theta)$ on multiple-qudit systems with an arbitrary $d$. 
Here, let us first introduce a virtual qudit in the correlation space in order to extend the notion of partner to multiple-qudit systems. 
A virtual qudit is characterized by a set of traceless Hermitian operators $\{\hat{T}_i\}_{i=1}^{d^2-1}$ satisfying
\begin{align}
 \hat{T}_i=\hat{V}^\dag \left(\hat{t}_i\otimes \mathbb{I}_d^{\otimes N-1}\right)\hat{V}\label{eq_algebra}
\end{align}
for a unitary operator $\hat{V}:\mathcal{H}\to\mathcal{H}$. Here, $\{\hat{t}_i\}_{i=1}^{d^2-1}$ is a basis of $\mathfrak{su}(d)$ algebra satisfying $\mathrm{Tr}_{\mathcal{H}_d}\left(\hat{t}_i\hat{t}_j\right)=d \delta_{ij}$ for $i,j=1,\cdots, d^2-1$. Assuming that the system is in a pure state $\ket{\Psi}$, the corresponding qudit-state $\hat{\rho}$ is defined as
\begin{align}
\hat{\rho}\equiv \frac{1}{d} \sum_{\mu=0}^{d^2-1}\Braket{\Psi|\hat{T}_\mu|\Psi}\hat{t}_{\mu},
\end{align}
where we have defined $\hat{T}_0=\mathbb{I}_d^{\otimes N}$ and $\hat{t}_0\equiv\mathbb{I}_d$. It can be shown that $\hat{\rho}$ is a unit trace positive-semidefinite operator, meaning that $\hat{\rho}$ defines a quantum state. The density operator $\hat{\rho}$ is characterized by the expectation values of generators $\Braket{\Psi|\hat{T}_{\mu}|\Psi}$, and the virtual qudit is defined in the correlation space. Quantum operations on the $N$-qudit system affect the expectation values, and achieve quantum operations on the virtual qudit. 

%Ambiguity of virtual-qudit operators
It should be noted that there is an ambiguity in the operators representing the same physical system. As an instructive example, let us consider a virtual qudit defined by
\begin{align}
 \hat{T}_i=\hat{t}_i\otimes \mathbb{I}_d^{\otimes N-1}
\end{align}
for the system in a pure state $\ket{\Psi}\in\mathcal{H}$. These operators characterize the first real qudit.
On the other hand, the state of a virtual qudit characterized by
\begin{align}
\hat{T}_i' \equiv e^{i\sum_{\mu=0}^{d^2-1}c_{\mu}\hat{T}_{\mu}}\hat{T}_{i}e^{-i\sum_{\mu=0}^{d^2-1}c_\mu\hat{T}_{\mu}}=\hat{t}'_{i}\otimes \mathbb{I}_d^{\otimes N-1}
\end{align}
with $c_\mu\in\mathbb{R}$ also characterize the same real qubit with different basis $\hat{t}'_i\equiv e^{i\sum_{\mu=0}^{d^2-1}c_{\mu}\hat{t}_{\mu}}\hat{t}_ie^{-i\sum_{\mu=0}^{d^2-1}c_{\mu}\hat{t}_{\mu}}$. In general, if two sets of operators $\left\{\hat{T}_i\right\}_{i=1}^{d^2-1}$ and $\left\{\hat{T}_i'\right\}_{i=1}^{d^2-1}$ are connected by a unitary operation generated by $\left\{\hat{T}_i\right\}_{i=1}^{d^2-1}$, they represents the same virtual qudit. In that case, we say $\left\{\hat{T}_i'\right\}_{i=1}^{d^2-1}$ is equivalent to $\left\{\hat{T}_i\right\}_{i=1}^{d^2-1}$. This ambiguity becomes significant when we discuss the non-uniqueness of QIC later.

%Definition of partner qudit
Let us define and identify the partner qudit.
For a given set of virtual-qudit operators $\left\{\hat{T}_i^{(A)}\right\}_{i=1}^{d^2-1}$, its partner qudit in a pure state $\ket{\Psi}$ is characterized by a set of traceless Hermitian operators $\left\{\hat{T}_i^{(B)}\right\}_{i=1}^{d^2-1}$ satisfying the following three conditions: (i) Algebra: it satisfies Eq.~(\ref{eq_algebra}), (ii) Locality: $\left[\hat{T}_i^{(A)},\hat{T}_j^{(B)}\right]=0$ for all $i,j=1,\cdots,d^2-1$, (iii) Purification: the operator $\hat{\rho}_{AB}\equiv \frac{1}{d^2}\sum_{\mu,\nu=0}^{d^2-1}\Braket{\Psi|\hat{T}_\mu^{(A)}\hat{T}_{\nu}^{(B)}|\Psi}\hat{t}_{\mu}\otimes\hat{t}_{\nu}$ represents a pure state in the correlation space. The partner qudit is identified by the following procedure: For a given $\hat{T}_i^{(A)}= \hat{V}^{(A)}{}^\dag\left(\hat{t}_i\otimes \mathbb{I}_d^{\otimes N-1}\right)\hat{V}^{(A)}$ and a pure state $\ket{\Psi}$, consider a pure state $\ket{\Psi'}\equiv \hat{V}^{(A)}\ket{\Psi}$. The Schmidt decomposition of the state is given by $\ket{\Psi'}=\sum_{i=1}^d \sqrt{p_i}\ket{\phi_i} \ket{\psi_i}$ with some probability distribution $\left\{p_i\right\}_{i=1}^d$ and orthonormal bases $\left\{\ket{\phi_i}\right\}_{i=1}^d$ and $\left\{\ket{\psi_i}\right\}_{i=1}^d$ on the Hilbert spaces $\mathcal{H}_d$ and $\mathcal{H}_d^{\otimes N-1}$, respectively. There always exists a unitary operator $\hat{v}:\mathcal{H}_d^{\otimes N-1}\to \mathcal{H}_d\otimes\mathcal{H}_d^{\otimes N-2}$ such that $\hat{v}\ket{\psi_i}=\ket{\psi_i'}\ket{\chi}$, where $\left\{\ket{\psi_i'}\right\}_{i=1}^d$ is an orthonormal basis of $\mathcal{H}_d$ and $\ket{\chi}$ is a unit vector in $\mathcal{H}_d^{\otimes N-2}$. Defining a unitary operator
\begin{align}
 \hat{V}^{(B)}\equiv \left(\sum_{i,j=1}^d\ket{\phi_i}\bra{\phi_j}\otimes \ket{\psi_j'}\bra{\psi_i'}\otimes \mathbb{I}_d^{\otimes N-2}\right)\left(\mathbb{I}_d\otimes \hat{v}\right)\hat{V}^{(A)},\label{eq_ub}
\end{align} 
the operators $\hat{T}_i^{(B)}=\hat{V}^{(B)}{^\dag}\left(\hat{t}_i\otimes \mathbb{I}^{\otimes N-1}\right)\hat{V}^{(B)}$ give the partner qudit. The corresponding state in the correlation space is given by $\hat{\rho}_{AB}=\ket{\Psi_{AB}}\bra{\Psi_{AB}}$, where $\ket{\Psi_{AB}}\equiv \sum_{i=1}^d\sqrt{p_i}\ket{\phi_i}\ket{\psi_i'}$ is a pure state for a virtual two-qudit system.

% Information extraction using partners
A pair of partners gives a way to extract the delocalized information. Suppose that information of unknown parameter $\theta$ is injected by a write operation $\hat{W}(\theta)=e^{-i\theta\hat{T}}$ on an $N$-qudit system in a pure state $\ket{\Psi}$. Here, $\hat{T}$ is assumed to satisfy $\hat{T}=\hat{U}^\dag\left(\hat{t}\otimes \mathbb{I}_d^{\otimes N-1}\right)\hat{U}$ for a unitary operator $\hat{U}$. Consider a pair of partner qudits $AB$ characterized by
\begin{align}
 \hat{T}_i^{(A)}\equiv \hat{U}^\dag\left(\hat{t_i}\otimes \mathbb{I}_d^{\otimes N-1}\right)\hat{U}, \quad\hat{T}_i^{(B)}=\hat{V}^{(B)} {}^\dag\left(\hat{t}_i\otimes \mathbb{I}_d^{\otimes N-1}\right)\hat{V}^{(B)},\label{eq_spartner}
\end{align}
where $\hat{V}^{(B)}$ is defined in Eq.~(\ref{eq_ub}) with $\hat{V}^{(A)}=\hat{U}$. Due to the write operation on the real $N$-qudit system, the correlation space state evolves into
\begin{align}
 \hat{\rho}_{AB}(\theta)
&\equiv \frac{1}{d^2}\sum_{\mu,\nu=1}^{d^2-1}\Braket{\Psi(\theta)|\hat{T}_{\mu}^{(A)}\hat{T}_{\nu}^{(B)}|\Psi(\theta)}\hat{t}_{\mu}\otimes \hat{t}_{\nu}\nonumber\\
&= \frac{1}{d^2}\sum_{\mu,\nu=1}^{d^2-1}\Braket{\Psi|e^{i\theta\hat{T}}\hat{T}_{\mu}^{(A)}e^{-i\theta\hat{T}}\hat{T}_{\nu}^{(B)}|\Psi}\hat{t}_{\mu}\otimes \hat{t}_{\nu}\nonumber\\
 &= \frac{1}{d^2}\sum_{\mu,\nu=1}^{d^2-1}\Braket{\Psi_{AB}|\left(e^{i\theta \hat{t}}\hat{t_\mu}e^{-i\theta\hat{t}}\otimes \mathbb{I}_d\right)\left(\mathbb{I}_d\otimes \hat{t}_\nu\right)|\Psi_{AB}}\hat{t}_{\mu}\otimes\hat{t}_{\nu}\nonumber\\
 &=\frac{1}{d^2}\sum_{\mu,\nu=1}^{d^2-1}\Braket{\Psi_{AB}|\hat{t_\mu}\otimes  \hat{t}_\nu|\Psi_{AB}}e^{-i\theta\hat{t}}\hat{t}_{\mu}e^{i\theta\hat{t}}\otimes\hat{t}_{\nu}\nonumber\\
 &=\left(\hat{w}(\theta)\otimes\mathbb{I}_d\right)\ket{\Psi_{AB}}\bra{\Psi_{AB}} \left(\hat{w}(\theta)\otimes\mathbb{I}_d\right)^\dag.
\end{align}
Therefore, the write operation $\hat{W}(\theta)$ achieves a write unitary operation $\hat{w}(\theta)\otimes\mathbb{I}_d$ on the partner $A$ in the correlation space. The initial state $\hat{\rho}_{AB}=\ket{\Psi_{AB}}\bra{\Psi_{AB}}$ and hence $\hat{\rho}_{AB}(\theta)$ are pure, implying that the information of $\theta$ is confined in the two-qudit state for the partners. The SWAP operation of a virtual qudit characterized by $\left\{\hat{T}_i\right\}_{i=1}^{d^2-1}$ is expressed as
\begin{align}
 \hat{U}_{\mathrm{swap}}\equiv\frac{1}{d}\sum_{\mu=0}^{d^2-1}\hat{T}_{\mu}\otimes \hat{t}_{\mu}:\mathcal{H}\otimes\mathcal{H}_d^{\mathrm{(ext.)}}\to\mathcal{H}\otimes\mathcal{H}_d^{\mathrm{(ext.)}},\label{eq_swap}
\end{align}
where $\hat{t}_\mu$ denotes the basis of $\mathfrak{su}(d)$ algebra for an external qudit system. The delocalized information can be perfectly retrieved by the SWAP operations with $\left\{\hat{T}_i^{(A)}\right\}_{i=1}^{d^2-1}$ and $\left\{\hat{T}_i^{(B)}\right\}_{i=1}^{d^2-1}$. 

%Different partners pictures
For a multiple-qudit system in a pure state $\ket{\Psi}$ and a write operation $\hat{W}(\theta)=e^{-i\theta\hat{T}}$, a pair of partners is characterized by operators defined in Eq.~(\ref{eq_spartner}).
For a unitary operator $\hat{V}$ which preserves $\hat{T}$, it is possible to introduce another pair of partners $A'B'$ as
\begin{align}
 \hat{T}_i{}^{(A')}=\hat{V}^\dag \hat{T}^{(A)}\hat{V},\quad \hat{T}_{i}^{(B')}=\hat{V}^\dag\hat{T}^{(B)}\hat{V}.
\end{align}
Since $\hat{T}$ is preserved, the write operation $\hat{W}(\theta)$ induces the same local write operation $\hat{w}(\theta)$ on the qudit $A'$ in the two-qudit system $A'B'$ in the correlation space. The amount of entanglement between $A'B'$ depends on the choice of the unitary operator $\hat{V}$ since the density matrix of the qudit $A'$ is given by
\begin{align}
 \hat{\rho}_{A'}&=\frac{1}{d}\sum_{\mu=0}^{d^2-1}\Braket{\Psi|\hat{T}_{\mu}^{(A')}|\Psi}\hat{t}_{\mu} =\frac{1}{d}\sum_{\mu=0}^{d^2-1}\Braket{\Psi_{V}|\hat{t}_{\mu}\otimes\mathbb{I}_d|\Psi_{V}}\hat{t}_{\mu},\label{eq_rhoA'}
\end{align}
where we have defined $\ket{\Psi_V}\equiv\hat{V}\ket{\Psi}$. The amount of entanglement is invariant under the write operation since it corresponds to a local unitary operation on the virtual qudit $A'$:
\begin{align}
 \hat{\rho}_{A'}(\theta)&\equiv\frac{1}{d}\sum_{\mu=0}^{d^2-1}\Braket{\Psi(\theta)|\hat{T}_{\mu}^{(A')}|\Psi(\theta)}\hat{t}_{\mu} =\frac{1}{d}\sum_{\mu=0}^{d^2-1}\Braket{\Psi_{V}|e^{i\theta\hat{t}}\hat{t}_{\mu}e^{-i\theta\hat{t}}\otimes\mathbb{I}_d|\Psi_{V}}\hat{t}_{\mu}\nonumber\\
 &=\hat{w}(\theta)\hat{\rho}_{A'}\hat{w}(\theta)^\dag.
\end{align}

%Maximally entangled partners
As an extreme case of partners, let us investigate whether maximally entangled partners exist. In this case, the reduced state of virtual qudits are invariant under the write operation, meaning that the whole information is stored in non-local correlations. As an example, consider a two-qudit system and fix the generator of write operation as $\hat{T}=\hat{t}_1\otimes \mathbb{I}_d$, where $\left\{\hat{t}_i\right\}_{i=1}^{d^2-1}$ denotes a basis of $\mathfrak{su}(d)$. The pair of partners $A'B'$ is maximally entangled if and only if the purity of $A'$ defined by $\mathrm{Tr}\left(\hat{\rho}_{A'}^2\right)=\frac{1}{d}$. Since
\begin{align}
 \mathrm{Tr}\left(\hat{\rho}_{A'}^2\right)=\frac{1}{d}\left(1+\sum_{i=1}^{d^2-1}\Braket{\Psi_V|\hat{t}_{i}\otimes \mathbb{I}_d|\Psi_V}^2\right)
\end{align}
holds, this condition is equivalent to
\begin{align}
 \Braket{\Psi_V|\hat{t}_{i}\otimes \mathbb{I}_d|\Psi_V}=0
\end{align}
for all $i=1,\cdots,d^2-1$. Since $\hat{V}$ preserves $\hat{t}_1\otimes \mathbb{I}_d$, it can be satisfied only if $\Braket{\Psi|\hat{t}_{1}\otimes \mathbb{I}_d|\Psi}=0$, which does not always hold. This example shows that maximally entangled partners do not exist in general. The fact that information can be hidden from the subsystems perfectly only in specific situations is consistent with the results in \cite{BP,MPSS}, though our setup is different from theirs.

%The definition and existence of QIC
The opposite situation gives the notion of QIC. 
A QIC is a virtual qudit in a pure state which perfectly confines the injected information of $\theta$. 
It is a non-trivial question whether a QIC exists, as is suggested from the fact that maximally entangled partners do not exist in general. 
Surprisingly, however, we can construct a QIC for arbitrary state and write operation. 
In order to show it, we use the eigenvalue decomposition $\hat{t}=\sum_{i=1}^{d}r_i\ket{\phi_i}\bra{\phi_i}$, where $r_i\in\mathbb{R}$ and $\left\{\ket{\phi_i}\right\}_{i=1}^d$ is an orthonormal basis. By using the basis, the state $\ket{\Psi'}\equiv \hat{U}\ket{\Psi}$ can be expanded as
\begin{align}
 \ket{\Psi'}=\sum_{i=1}^dc_i\ket{\phi_i}\otimes\ket{\psi_i},
\end{align}
where $c_i\in\mathbb{C}$ and $\ket{\psi_i}\in\mathcal{H}^{\otimes N-1}$ are unit vectors, which are not orthogonal to each other in general. Let us define a unitary operator
\begin{align}
 \hat{V}\equiv \exp{\left(-i\left(\sum_{i=1}^d\ket{\phi_i}\bra{\phi_i}\otimes \hat{h}_i\right)\right)},\label{eq_qicv}
\end{align}
where $\left\{h_i\right\}_{i=1}^{d}$ are Hermitian operators satisfying $e^{-ih_i}\ket{\psi_i}=\ket{\psi}$ for a reference unit vector $\ket{\psi}\in\mathcal{H}^{\otimes N-1}$. This transformation preserves $\hat{t}\otimes \mathbb{I}_d$. Defining $\ket{\Phi}\equiv \sum_{i=1}^d c_i\ket{\phi_i}$, we get
\begin{align}
 \hat{V}\ket{\Psi'}=\sum_{i=1}^dc_i\ket{\phi_i}\otimes e^{-ih_i}\ket{\psi_i}=\ket{\Phi}\otimes\ket{\psi},\label{eq_qicstate}
\end{align}
meaning that the virtual qudit characterized by a set of operators $\left\{\hat{T}_i^{(\mathrm{QIC})}\equiv \hat{U}^\dag \hat{V}^\dag\left(\hat{t}_i\otimes\mathbb{I}_d^{\otimes N-1}\right)\hat{V}\hat{U}\right\}_{i=1}^{d^2-1}$ is in a pure state $\hat{\rho}_{\mathrm{QIC}}=\ket{\Phi}\bra{\Phi}$. Therefore, this virtual qudit is a QIC. 
By using the QIC operators $\left\{\hat{T}_i^{(\mathrm{QIC})}\right\}_{i=1}^{d^2-1}$, the whole information can be perfectly retrieved by the SWAP operation given in Eq.~(\ref{eq_swap}).

%The non-uniqueness of QIC
In the above proof, unitary operators $\left\{e^{-i\hat{h}_i}\right\}_{i=1}^d$ are arbitrarily chosen as long as it satisfies $e^{-i\hat{h}_i}\ket{\psi_i}=\ket{\psi}$. The non-uniqueness of QIC is shown from this fact. As an example, consider a one-parameter family of unitary operator
\begin{align}
 \hat{V}(r)= e^{-i r\hat{t}\otimes \ket{\psi}\bra{\psi}}\hat{V}=\exp{\left(-i\left(\sum_{i=1}^d\ket{\phi_i}\bra{\phi_i}\otimes \left(\hat{h}_i+r r_i\ket{\psi}\bra{\psi}\right)\right)\right)},
\end{align}
where $r\in\mathbb{R}$.
The operators defined by
\begin{align}
 \hat{T}^{(\mathrm{QIC})}_i(r)\equiv \hat{U}^\dag\hat{V}(r)^\dag\left(\hat{t}_i\otimes \mathbb{I}_d^{\otimes N-1}\right)\hat{V}(r)\hat{U}
\end{align}
also characterize a QIC. Since
\begin{align}
 \hat{T}_i^{(\mathrm{QIC})}(r)=\hat{U}^\dag\hat{V}^\dag\left(\hat{t}_i\otimes \mathbb{I}_{d}^{\otimes N-1}+\left(e^{ir\hat{t}}\hat{t}_ie^{-ir\hat{t}}-\hat{t}_i\right)\otimes \ket{\psi}\bra{\psi}\right)\hat{V}\hat{U}\label{eq_noneq} 
\end{align}
holds, there exists $r^*$ such that $\left\{\hat{T}_i(r^{*})\right\}_{i=1}^{d^2-1}$ is not equivalent to $\left\{\hat{T}_i\right\}_{i=1}^{d^2-1}$. More generally, by using an Hermitian operator $\hat{h}$ satisfying $\hat{h}\ket{\psi}=r\ket{\psi}$ with a real number $r$, a set of QIC operators are constructed as
\begin{align}
 \hat{T}_i^{(\mathrm{QIC})}(\hat{h})
&\equiv\hat{U}^\dag\hat{V}(\hat{h})^\dag\left(\hat{t}_i\otimes \mathbb{I}_d\right)\hat{V}(\hat{h})\hat{U}
\end{align}
where $ \hat{V}(\hat{h})\equiv e^{-i\hat{t}\otimes \hat{h}}\hat{V}$.
This implies that for any state and write operation, there exist different QICs. In \cite{YWH}, the non-uniqueness of QIC has been proven only for $d=2$. The example of the Greenberger--Horne--Zeilinger state presented there clearly shows that different QICs gives different ways to process the injected information. 

%Natural choice of QIC operators
It should be noted that even when information is localized, QIC is non-unique. For example, let us consider a two-qudit system in a product state $\ket{\Psi}=\ket{\phi}\ket{\psi}$. If we inject the information by a write operator $\hat{W}(\theta)=\hat{w}(\theta)\otimes \mathbb{I}_d$, the information is localized in the first qudit and a set of QIC operators are given by $\hat{T}_i=\hat{t}_i\otimes \mathbb{I}_d$. From Eq.~(\ref{eq_noneq}), the operators 
\begin{align}
 \hat{T}_i(r)=\hat{T}_i+\left(e^{ir\hat{t}}\hat{t}_ie^{-ir\hat{t}}-\hat{t}_i\right)\otimes \ket{\psi}\bra{\psi}
\end{align}
also characterize another QIC. In this case, one would feel that the set of operators $\left\{\hat{T}_i\right\}_{i=1}^{d^2-1}$ is more natural and physically significant than that with $r\neq0$ since the former one agrees with the observation that information is spatially localized. It remains an open question whether there is a useful criterion to single out a important or convenient set of operators in general setup. For shift write operations on Gaussian states, there is a criterion with which a QIC is uniquely selected, as we will see in the next section.

Before concluding this section, we briefly discuss multi-parameter write operations. A QIC in a multiple-qudit system is capable of $(d-1)$ independent parameters. It should be noted that multi-parameter information storage can be achieved for $d\geq 3$. Let $\left\{\hat{c_i}\right\}_{i=1}^{d-1}$ be a set of commutative operators in $\mathfrak{su}(d)$. Suppose that $(d-1)$ real unknown parameters $\left\{\theta_i\right\}_{i=1}^{d-1}$ are injected on a pure state $\ket{\Psi}$ by write operations
\begin{align}
 \hat{W}_i(\theta_i)\equiv e^{-i\theta_i\hat{C}_i}=\hat{U}^\dag\left(e^{-i\theta_i\hat{c}_i}\otimes \mathbb{I}_d^{\otimes N-1}\right)\hat{U},
\end{align}
where we have defined $\hat{C}_i\equiv \hat{U}^\dag \left(\hat{c}_i\otimes \mathbb{I}_d^{\otimes N-1}\right)\hat{U}$. The state after operations are given by
\begin{align}
 \ket{\Psi(\bm{\theta})}\equiv\hat{W}_1(\theta_1)\cdots\hat{W}_{d-1}(\theta_{d-1})\ket{\Psi},
\end{align}
where we have introduced $\bm{\theta}\equiv(\theta_1,\cdots,\theta_{d-1})$.
As a quantifier of precision of multi-parameter estimation, the symmetric logarithmic derivative (SLD) Fisher information matrix $F_{\bm{\theta}}$ is often adopted \cite{QF}. Its element is defined as
\begin{align}
 \left(F_{\bm{\theta}}\right)_{ij}\equiv \Braket{\Psi(\bm{\theta})|\frac{1}{2}\left(\hat{L}_i\hat{L}_j+\hat{L}_j\hat{L}_i\right)|\Psi(\bm{\theta})},
\end{align}
where $\hat{L}_i$ is a solution of the following equation with $\hat{\rho}_{\bm{\theta}}\equiv \ket{\Psi(\bm{\theta})}\Bra{\Psi(\bm{\theta})}$:
\begin{align}
 \partial_{\theta_i}\hat{\rho}_{\bm{\theta}}=\frac{1}{2}\left(\hat{\rho}_{\bm{\theta}}\hat{L}_i+\hat{L}\hat{\rho}_{\bm{\theta}}\right).
\end{align}
The operators $\hat{L}_i$ are called SLD operators. 
In our setup, SLD operators are easily obtained as
\begin{align}
 \hat{L}_i=2i\left(\hat{\rho}_{\bm{\theta}} \hat{C}_i-\hat{C}_i\hat{\rho}_{\bm{\theta}}\right).
\end{align}
From this equation, the Fisher information matrix is given by
\begin{align}
 \left(F_{\bm{\theta}}\right)_{ij}=4\Braket{\Psi|\Delta\hat{C}_i\Delta\hat{C}_j|\Psi},\label{eq_QFI}
\end{align}
where we have defined $\Delta\hat{C}_i\equiv \hat{C}_i -\Braket{\Psi|\hat{C}_i|\Psi}$. If the Fisher matrix is non-degenerate, parameters are independently imprinted. A QIC for this series of write operations can be constructed in the following way: Since $\hat{c}_i$ commutes with each other, they are simultaneously diagonalizable. That is, there exists an orthonormal basis $\left\{\ket{\phi_i}\right\}_{i=1}^{d}$ such that 
\begin{align}
 \hat{c}_i=\sum_{j=1}^d r_{ij}\ket{\phi_j}\bra{\phi_j}
\end{align}
for some real numbers $r_{ij}$. From the same argument as in Eqs.~(\ref{eq_qicv}) and (\ref{eq_qicstate}), the QIC operators are obtained. The whole information of $(d-1)$ parameters is simultaneously extracted by the SWAP operation given in Eq.~(\ref{eq_swap}). 

There is another setup of multi-parameter information storage by using commutative write operations. As an example, consider two write operations $\hat{W}_1(\theta_1)\equiv e^{-i\theta_1\hat{T}_1}$ and $\hat{W}_2(\theta_2)\equiv e^{-i\theta_2\hat{T}_2}$, where $\hat{T}_1\equiv \hat{t}\otimes \mathbb{I}_d^{\otimes N-1}$ and $\hat{T}_2\equiv \mathbb{I}_d\otimes \hat{s}\otimes \mathbb{I}_d^{\otimes N-2}$ with $\hat{t},\hat{s}\in\mathfrak{su}(d)$. The commutativity of the write operations does not imply that the information can be retrieved independently. In other words, sets of QIC operators for two write operations may not commute with each other. In the case of multiple-qudit systems, it is difficult to discuss the commutativity of QIC operators since we do not have a simple formula to construct the QIC operators. However, in the case of write shift operations on CV systems in Gaussian state, it is possible to derive the condition under which multi-parameter information can independently be retrieved as we will see in the next section.

\section{QIC in CV systems}\label{sec_inf_dim}
In the previous section, we have shown that a QIC always exists for the information injected into multiple-qudit systems by write operations in the form of $\hat{W}(\theta)=e^{-i\theta \hat{U}\left(\hat{t}\otimes \mathbb{I}_d\right)\hat{U}}$. 
Here, we extend the arguments to CV systems: multiple-harmonic oscillator (HO) systems and the scalar field. 

\subsection{QIC in multiple-HO systems}
Let us consider an $N$-HO system whose canonical variables are denoted by $\hat{\bm{r}}\equiv (\hat{q}_1,\hat{p}_1,\cdots,\hat{q}_N,\hat{p}_N)\sps{T}$. The canonical commutation relations are summarized as
\begin{align}
 \left[\hat{\bm{r}},\hat{\bm{r}}\sps{T}\right]=i\Omega,
\end{align}
where
\begin{align}
 \Omega\equiv \bigoplus_{n=1}^N
\begin{pmatrix}
 0&1\\
 -1&0 
\end{pmatrix}
\end{align}
is an anti-symmetric matrix. Let us inject information of $\theta$ by a write operation $\hat{W}(\theta)=e^{-i\theta\hat{Q}}$ on a pure state $\ket{\Psi}$, where $\hat{Q}$ satisfies
\begin{align}
 \hat{Q}=\hat{U}^\dag\left(\hat{q}_1\otimes \mathbb{I}^{\otimes N-1}\right)\hat{U}\label{eq_qnonlinear}
\end{align}
for some unitary operator $\hat{U}$, where $\mathbb{I}$ is the identity operator for a single HO system. For notational simplicity, we will omit $\mathbb{I}$ and $\otimes$ when there is no risk of confusion.
Defining $\hat{P}\equiv \hat{U}^\dag\hat{p}_1\hat{U}$, the canonical commutation relation is satisfied $\left[\hat{Q},\hat{P}\right]=i$, implying that the set of operators $\left(\hat{Q},\hat{P}\right)$ characterizes a mode. For a given mode and an arbitrary pure state $\ket{\Psi}$, it is possible to construct its purification partner \cite{TYH}. Since the composite subsystem of the modes defined by $\left(\hat{Q},\hat{P}\right)$ and its partner is in a pure state, the information is perfectly confined in this subsystem. Therefore, the picture of partners storing information is valid for $N$-HO systems. 

What about the picture of QIC?
We can extend the proof of the existence of a QIC to the $N$-HO systems in the following way: The eigenvalue decomposition of the operator $\hat{q}_1$ is given by
\begin{align}
 \hat{q}_1=\int_{-\infty}^\infty dq\, q \ket{q}\bra{q}.
\end{align}
Expanding the state $\ket{\Psi'}\equiv \hat{U}\ket{\Psi}$ as
\begin{align}
 \ket{\Psi'}=\int_{-\infty}^\infty dq\, c(q)\ket{q}\otimes \ket{\psi_q},
\end{align}
we can introduce the unitary operator as
\begin{align}
 \hat{V}\equiv \exp{\left(-i\int_{-\infty}^\infty dq \, \ket{q}\bra{q}\otimes \hat{h}_q\right)},\label{eq_v_ho}
\end{align}
where $\hat{h}_q$ satisfies the condition that $e^{-i\hat{h}_q}\ket{\psi_q}=\ket{\psi}$ for a reference state $\ket{\psi}$ of $(N-1)$-HO system. Since $\hat{V}$ preserves $\hat{q}_1$ and satisfies
\begin{align}
 \hat{V}\hat{U}\ket{\Psi}=\left(\int_{-\infty}^{\infty}dq \, c(q)\ket{q}\right)\otimes \ket{\psi},
\end{align} 
the set of operators
\begin{align}
 \hat{Q}^{(\mathrm{QIC})}&\equiv \hat{U}^\dag \hat{V}^\dag \hat{q}_1 \hat{V}\hat{U}=\hat{Q},\\
 \hat{P}^{(\mathrm{QIC})}&\equiv \hat{U}^\dag \hat{V}^\dag \hat{p}_1 \hat{V}\hat{U}
\end{align}
characterizes a QIC. Therefore, as long as the unitary operator defined in Eq.~(\ref{eq_v_ho}) is well defined, we can construct a QIC. 

Hereafter, we mainly investigate the information injected into pure Gaussian states by write operations generated by linear combination of the canonical operators. That is, we restrict ourselves in the case when the write operation is given by
\begin{align}
 \hat{W}(\theta)=e^{-i\theta\hat{Q}},\quad \hat{Q}\equiv \bm{v}^{\mathrm{T}}\hat{\bm{r}}\label{eq_qlinear}
\end{align}
where $\bm{v}\in\mathbb{R}^{2N}$. The coefficients $\bm{v}$ correspond to the weighting functions \cite{TYH} that characterize the interaction between the $N$-HO system and a device that detects a mode. In the case of the scalar field theory, such devices are called Unruh-de Witt detector \cite{U,dW}. The operator $\hat{Q}$ in the form in Eq.~(\ref{eq_qlinear}) is a special example of operators defined in Eq.~(\ref{eq_qnonlinear}). This write operations shift the first moment of states as follows:
\begin{align}
 \Braket{\Psi(\theta)|\hat{\bm{r}}|\Psi(\theta)}= \Braket{\Psi|\hat{\bm{r}}|\Psi}+\theta\Omega\bm{v}.
\end{align}
Therefore, we call them shift write operations. 

A pure Gaussian state $\ket{\Psi}$ is fully characterized by its first and second moments of the canonical variables. The second moments are summarized by the covariance matrix
\begin{align}
 M\equiv\mathrm{Re}\left(\Braket{\Psi|\hat{\bm{R}}\hat{\bm{R}}^{\mathrm{T}}|\Psi}\right),
\end{align}
where we have defined
\begin{align}
 \hat{\bm{R}}\equiv\hat{\bm{r}}-\Braket{\Psi|\hat{\bm{r}}|\Psi}
\end{align}
so that $\Braket{\Psi|\hat{\bm{R}}|\Psi}=\bm{0}$ holds. Our task is to find an operator $\hat{P}$ such that $\left[\hat{Q},\hat{P}\right]=i$ holds and the mode characterized by $\left(\hat{Q},\hat{P}\right)$ is in a pure state for the given state $\ket{\Psi}$. If we could find it, after the write operation $\hat{W}(\theta)=e^{-i\theta\hat{Q}}$, the mode remains in a pure state, implying that the mode is a QIC. We call such an operator $\hat{P}$ the conjugate QIC operator of the operator $\hat{Q}$. Taking into account that the purity of Gaussian state is characterized by the second moments, we pose the following ansatz:
\begin{align}
 \hat{P}=\bm{u}^{\mathrm{T}}\hat{\bm{R}}=\bm{u}^{\mathrm{T}}\hat{\bm{r}}-\bm{u}^{\mathrm{T}}\Braket{\Psi|\hat{\bm{r}}|\Psi},\label{eq_p_shift}
\end{align}
where $ \bm{u}\in\mathbb{R}$ must satisfy
\begin{align}
 \bm{v}^{\mathrm{T}}\Omega\bm{u}=1\label{eq_vOmegau}
\end{align}
so that $\left[\hat{Q},\hat{P}\right]=i$ holds. The second term in Eq.~(\ref{eq_p_shift}) is a constant factor which becomes significant when we study multi-parameter write operations as we will see later.

For a mode characterized by $\left(\hat{Q},\hat{P}\right)$, its covariance matrix is defined by
\begin{align}
 m\equiv 
\begin{pmatrix}
 \mathrm{Re}\left(\Braket{\Psi|\hat{Q}^2|\Psi}\right)&\mathrm{Re}\left(\Braket{\Psi|\hat{Q}\hat{P}|\Psi}\right)\\
 \mathrm{Re}\left(\Braket{\Psi|\hat{P}\hat{Q}|\Psi}\right)& \mathrm{Re}\left(\Braket{\Psi|\hat{P}^2|\Psi}\right)
\end{pmatrix}=
\begin{pmatrix}
 \bm{v}\sps{T}M\bm{v}&\bm{v}\sps{T}M\bm{u}\\
 \bm{u}\sps{T}M\bm{v}&\bm{u}\sps{T}M\bm{u} 
\end{pmatrix},
\end{align}
where we have re-defined $\hat{Q}$ as $\hat{Q}=\bm{v}^{\mathrm{T}}\hat{\bm{R}}$. The difference between $\hat{Q}=\bm{v}^{\mathrm{T}}\hat{\bm{r}}$ and $\hat{Q}=\bm{v}^{\mathrm{T}}\hat{\bm{R}}$ in the write operation $\hat{W}(\theta)=e^{-\theta\hat{Q}}$ results in an unimportant global phase rotation. 
The entanglement entropy between the mode and its complement system is given by \cite{EE}
\begin{align}
 S_{\mathrm{EE}}=\sqrt{1+g^2}\ln{\left(\frac{1}{g}\left(\sqrt{1+g^2}+1\right)\right)}+\ln{\left(\frac{g}{2}\right)},
\end{align}
where $g$ is defined by
\begin{align}
 g\equiv \sqrt{4\det{m}-1}.
\end{align} 
The entanglement entropy becomes zero if and only if $g=0$. It should be noted that $\det{m}\geq\frac{1}{4}$ is always satisfied because of the uncertainty relationship.

Similar to the case of virtual qudits, we also have the ambiguity of the choice in the operators representing a mode. That is, for a unitary operator $\hat{U}$ generated by $\left(\hat{Q},\hat{P}\right)$, the set of operators $\left(\hat{Q}',\hat{P}'\right)\equiv\left(\hat{U}^\dag\hat{Q}\hat{U},\hat{U}^\dag \hat{P}\hat{U}\right)$ also characterize the same mode. Taking
\begin{align}
 \hat{U}\equiv e^{-i\frac{\mathrm{Re}\left(\Braket{\hat{Q}\hat{P}}\right)}{2\Braket{\Psi|\hat{Q}^2|\Psi}}\hat{Q}^2},
\end{align}
$\hat{Q}'=\hat{Q}$ holds and $\hat{P}'$ satisfies
\begin{align}
 \mathrm{Re}\left(\Braket{\Psi|\hat{Q}'\hat{P}'|\Psi}\right)= \mathrm{Re}\left(\Braket{\Psi|\hat{P}'\hat{Q}'|\Psi}\right)=0\label{eq_vMu}.
\end{align}
To eliminate the ambiguity in the operators representing a QIC, we impose this condition. 
In terms of vectors $\bm{v}$ and $\bm{u}$, the above constraint is equivalent to
\begin{align}
 \bm{v}^{\mathrm{T}}M\bm{u}=0.
\end{align}

To summarize, the real vector $\bm{u}$ that determines $\hat{P}$ is the vector which minimize $\det{m}$ under the constraints Eqs.~(\ref{eq_vOmegau}) and (\ref{eq_vMu}). Thus, let us define a function
\begin{align}
 f(\bm{u})\equiv \left(\bm{v}^{\mathrm{T}}M\bm{v}\right)\left(\bm{u}^{\mathrm{T}}M\bm{u}\right)- \mu_1\left(\bm{v}\sps{T}\Omega\bm{u}-1\right)-\mu_2\left(\bm{v}\sps{T}M\bm{u}-0\right),\label{eq_lagrange}
\end{align}
where $\mu_1,\mu_2$ are Lagrange multipliers. Since
\begin{align}
 \frac{\partial}{\partial u_i}f(\bm{u})=2\left(\bm{v}\sps{T}M\bm{v}\right)\sum_{j=1}^{2N}M_{ij}u_j-\mu_1\sum_{j=1}^{2N}v_j\Omega_{ji}-\mu_2\sum_{j=1}^{2N}v_jM_{ji}
\end{align}
holds, the unique solution of $\frac{\partial}{\partial\bm{u}}f(\bm{u})=\bm{0}$ is given by
\begin{align}
 \bm{u}=-\frac{1}{\bm{v}\sps{T}M\bm{v}}\Omega M\bm{v}\label{eq_u}
\end{align}
with $\mu_1=\frac{1}{2}$ and $\mu_2=0$. Here, we have used the fact that 
\begin{align}
 M\Omega M =\frac{1}{4}\Omega\label{eq_mOmegam_dis}
\end{align}
holds for pure Gaussian states $\ket{\Psi}$ since there always exist a symplectic matrix $S$ such that $M=\frac{1}{2}S^{\mathrm{T}}S$ \cite{QCV}. From a straightforward calculation, it can be checked that for a given operator $\hat{Q}=\bm{v}\sps{T}\hat{\bm{R}}$ and a given Gaussian state $\ket{\Psi}$, the operator
\begin{align}
 \hat{P}=\frac{1}{\Braket{\Psi|\hat{Q}^2|\Psi}}\bm{v}\sps{T}M\Omega\hat{\bm{R}}\label{eq_conjQIC}
\end{align}
actually satisfies $\left[\hat{Q},\hat{P}\right]=i$ and $\det{m}=\frac{1}{4}$, implying that the mode defined by $\left(\hat{Q},\hat{P}\right)$ is in a pure state. Therefore, $\hat{P}$ is the conjugate QIC operator of $\hat{Q}$. It should be noted that the solution of Eq.~(\ref{eq_lagrange}) was unique. The ansatz given in Eq.~(\ref{eq_p_shift}) singles out the unique QIC. 

By using the operators $\left(\hat{Q},\hat{P}\right)$ characterizing the QIC, the information of $\theta$ can be retrieved by the following SWAP operation:
\begin{align}
 \hat{U}_{\mathrm{swap}}=\exp{\left(i\frac{\pi}{2}\left(\hat{Q}\hat{p}^{\mathrm{(ext.)}}-\hat{P}\hat{q}^{\mathrm{(ext.)}}\right)\right)},
\end{align}
where $\left(\hat{q}^{\mathrm{(ext.)}},\hat{p}^{\mathrm{(ext.)}}\right)$ is a set of canonical variables for an external device. Since the QIC operators derived here are linear combinations of canonical variables, this SWAP operation can be achieved by a bi-linear coupling between the system and the external device. 

Similar to the case of multiple-qudit systems, a QIC in CV systems can store multiple parameters. Let $\left(\hat{Q},\hat{P}\right)$ be a set of operators characterizing a QIC. Since they are linear combinations of canonical variables satisfying $\left[\hat{Q},\hat{P}\right]=i$, there exists a unitary operator $\hat{U}'$ such that $\hat{Q}=\hat{U}^\dag \hat{q}_1\hat{U}$ and $\hat{P}=\hat{U}^\dag \hat{p}_1\hat{U}$. Defining
\begin{align}
 \hat{T}_n\equiv \hat{U}^\dag \left(\ket{n}\bra{n}\otimes \mathbb{I}^{\otimes N-1}\right)\hat{U}
\end{align}
for non-negative integers $n$, they are Hermitian operators commute with each other. Here, $\left\{\ket{n}\right\}_{n=0}^\infty$ represents the set of eigenstates of the number operators of the first HO. 
The operator $\hat{T}_n$ can be also written as
\begin{align}
\hat{T}_n\equiv  \frac{1}{n!}\lim_{\beta\to\infty}e^{n\beta} \left(\frac{\partial}{\partial(-\beta)}\right)^ne^{-\beta\hat{a}\hat{a}^\dag},
\end{align}
where we have defined creation and annihilation operators of the QIC mode as 
\begin{align}
 \hat{a}^\dag \equiv \frac{1}{\sqrt{2}}\left(\hat{Q}-i\hat{P}\right),\quad \hat{a}\equiv \frac{1}{\sqrt{2}}\left(\hat{Q}+i\hat{P}\right).
\end{align}
By using a series of commutative write operation $\hat{W}_n(\theta_n)\equiv e^{-i\theta_n\hat{T}_n}$, a countably infinite number of parameters is stored independently in the single QIC mode.

Now, let us investigate another kind of multi-parameter write operations. Consider a series of shift write operations $\hat{W}_{i}(\theta_i)\equiv e^{-i\theta_i\hat{Q}_i}$, where $i=1,\cdots,k$ with an positive integer $k$, and $\hat{Q}_{i}\equiv \bm{v}_i^{\mathrm{T}}\hat{\bm{r}}$ with real vectors $\bm{v}_i\in\mathbb{R}$. Let us impose
\begin{align}
 \left[\hat{Q}_i,\hat{Q}_j\right]=i\bm{v}_i^{\mathrm{T}}\Omega\bm{v}_j=0\quad \text{for }1\leq i \leq j\leq  k\label{eq_QICcond1}
\end{align}
so that the write operation commute with each other. After the series of write operations, a Gaussian state $\ket{\Psi}$ evolves into
\begin{align}
 \ket{\Psi(\bm{\theta})}\equiv \hat{W}_1(\theta_1)\cdots \hat{W}_k(\theta_k)\ket{\Psi},
\end{align}
where we have defined $\bm{\theta}\equiv \left(\theta_1,\cdots,\theta_k\right)$. Now, suppose that we try to retrieve the information $\theta_1$ from $\ket{\Psi(\bm{\theta})}$. From Eq.~(\ref{eq_conjQIC}), the conjugate operator $\hat{P}_1$ to $\hat{Q}_1$ is given by
\begin{align}
 \hat{P}_1=\frac{1}{\Braket{\Psi|\hat{Q}_1^2|\Psi}}\bm{v}_1^{\mathrm{T}}M\Omega\hat{\bm{R}}_{2,\cdots,k},
\end{align}
where
\begin{align}
 \hat{\bm{R}}_{2,\cdots,k}\equiv \hat{\bm{R}}-\sum_{i=2}^{k}\theta_i\Omega\bm{v}_i
\end{align}
is a set of canonical variables whose first moments vanish for $\hat{W}_2(\theta_2)\cdots\hat{W}_k(\theta_k)\ket{\Psi}$. This means that for shift write operations, the conjugate QIC operator depends on unknown variables in general even when the write operations commute with each other. That is, if we want to extract the information of $\theta_1$, we need to know $\theta_2,\cdots,\theta_k$ unless $\bm{v}_1^{\mathrm{T}}M\bm{v}_j=0$ for all $j\neq 1$. Therefore, the information on each parameters are independently retrieved if and only if
\begin{align}
 \bm{v}_i^{\mathrm{T}}M\bm{v}_j=0\quad \text{for } 1\leq i< j\leq k \label{eq_QICcond2}
\end{align}
holds. 
Under this condition, the conjugate operators become independent of unknown parameters and are given by
\begin{align}
 \hat{P}_i=\frac{1}{\Braket{\Psi|\hat{Q}_i^2|\Psi}}\bm{v}_i^{\mathrm{T}}M\Omega \hat{\bm{R}}.
\end{align}
The condition given in Eq.~(\ref{eq_QICcond2}) ensures the commutativity of QIC operators:
\begin{align}
 \left[\hat{Q}_i,\hat{P}_j\right]=i\frac{\bm{v}_i^{\mathrm{T}}M\bm{v}_j}{\Braket{\Psi|\hat{Q}_1^2|\Psi}}=i\delta_{ij}.
\end{align}
The independence of parameters can be also checked from the Fisher matrix. In the same calculation as is done in Eq.~(\ref{eq_QFI}), the Fisher matrix is given by
\begin{align}
 \left(F_{\bm{\theta}}\right)_{ij}=4\Braket{\Psi|\hat{Q}_i^2|\Psi}\delta_{ij},
\end{align}
implying that it cannot be degenerated. 

From the viewpoint of unitary evolution, the conditions given in Eqs.~(\ref{eq_QICcond1}) and (\ref{eq_QICcond2}) ensures the invariance of the QIC operators under the write operations of other parameters. Suppose that information of $\theta_1$ is first injected by $\hat{W}_1(\theta_1)=e^{-i\theta_1\hat{Q}_1}$ on the $N$-HO system in a pure state $\ket{\Psi}$. The QIC operators are given by
\begin{align}
 \hat{Q}_1=\bm{v}_1^{\mathrm{T}}\hat{\bm{R}},\quad \hat{P}_1=\frac{1}{\Braket{\Psi|\hat{Q}_1^2|\Psi}}\bm{v}_1^{\mathrm{T}}M\Omega \hat{\bm{R}}.
\end{align}
As is discussed in \cite{YWH}, the time-evolution of the QIC operator is given by $\left(\hat{Q},\hat{P}\right)\mapsto \left(\hat{U}\hat{Q}\hat{U}^\dag,\hat{U}\hat{P}\hat{U}^\dag\right)$ when the system evolves as $\hat{\Psi}\mapsto \hat{U}\ket{\Psi}$ under a unitary operator $\hat{U}$. When we inject information of $\theta_2$ by $\hat{W}_2(\theta_2)=e^{-i\theta_2\hat{Q}_2}$, the QIC operators evolves into
\begin{align}
 \hat{W}_2(\theta_2)\hat{Q}_1\hat{W}_2(\theta_2)^\dag&=\bm{v}_1^{\mathrm{T}}\left(\hat{\bm{R}}-\theta_2\Omega\bm{v}_2\right)=\hat{Q}_1-\theta_2\bm{v}_1^{\mathrm{T}}\Omega\bm{v}_2,\\
 \hat{W}_2(\theta_2)\hat{P}_1\hat{W}_2(\theta_2)^\dag&=\frac{1}{\Braket{\Psi|\hat{Q}_1^2|\Psi}}\bm{v}_1^{\mathrm{T}}M\Omega \left(\hat{\bm{R}}-\theta_2\Omega\bm{v}_2\right)=\hat{P}_1+\frac{1}{\Braket{\Psi|\hat{Q}_1^2|\Psi}}\theta_2\bm{v}_1^{\mathrm{T}}M\bm{v}_2.
\end{align}
Therefore, if Eqs.~(\ref{eq_QICcond1}) and (\ref{eq_QICcond2}) are satisfied, the QIC operators $\left(\hat{Q}_1,\hat{P}_1\right)$ is invariant under the write operations of $\hat{W}_2(\theta_2)$.

\subsection{QIC in scalar field theory}
Let us consider a scalar field theory in $(d+1)$-dimensional spacetime. The scalar field $\hat{\phi}(\bm{x})$ and its conjugate momentum $\hat{\Pi}(\bm{x})$ satisfy the canonical commutation relation:
\begin{align}
\left[\hat{\phi}(\bm{x}),\hat{\phi}(\bm{y})\right]=0,\quad \left[\hat{\phi}(\bm{x}),\hat{\Pi}(\bm{y})\right]=i\delta^{(d)}(\bm{x}-\bm{y}),\quad \left[\hat{\Pi}(\bm{x}),\hat{\Pi}(\bm{y})\right]=0
\end{align}
for $\bm{x},\bm{y}\in\mathbb{R}^{d}$. Defining $\hat{\gamma}(\bm{x})\equiv \left(\hat{\phi}(\bm{x}),\hat{\Pi}(\bm{x})\right)^{\mathrm{T}}$, it can be summarized as $\left[\hat{\gamma}(\bm{x}),\hat{\gamma}(\bm{y})^{\mathrm{T}}\right]=i\Omega(\bm{x},\bm{y})$, where
\begin{align}
 \Omega(\bm{x},\bm{y})\equiv
\begin{pmatrix}
 0 &\delta^{(d)}(\bm{x}-\bm{y})\\
 -\delta^{(d)}(\bm{x}-\bm{y})& 0
\end{pmatrix}.
\end{align}
When the system is in a pure Gaussian state $\ket{\Psi}$, let us re-define the field and conjugate momentum as $\hat{\Gamma}(\bm{x})\equiv \hat{\gamma}(\bm{x})-\Braket{\Psi|\hat{\gamma}(\bm{x})|\Psi}$ so that the first moments vanish: $\Braket{\Psi|\hat{\Gamma}(\bm{x})|\Psi}=0$. The covariance matrix is defined as
\begin{align}
 M(\bm{x},\bm{y})\equiv \mathrm{Re}\left(\Braket{\Psi|\hat{\Gamma}(\bm{x})\hat{\Gamma}(\bm{y})^{\mathrm{T}}|\Psi}\right).
\end{align}

Now, let us define a shift write operation $\hat{W}(\theta)\equiv e^{-i\theta\hat{Q}}$ where 
\begin{align}
 \hat{Q}\equiv \int d^d\bm{x}\, v^{\mathrm{T}}(\bm{x})\hat{\Gamma}(\bm{x}),
\end{align}
and 
\begin{align}
  v(\bm{x})\equiv \left(v_1(\bm{x}),v_2(\bm{x})\right)^{\mathrm{T}}
\end{align}
are weighting functions.
As long as the continuum limit can be taken properly, the results in the previous subsection are applicable here. Therefore, the conjugate QIC operator $\hat{P}$ can be constructed as
\begin{align}
 \hat{P}=\frac{1}{\Braket{\Psi|\hat{Q}^2|\Psi}} \int d^d \bm{x}d^d\bm{y}d^d\bm{z} \, v(\bm{x})^{\mathrm{T}}M(\bm{x},\bm{y})\Omega(\bm{y},\bm{z})\hat{\Gamma}(\bm{z})\label{eq_conjP_field}
\end{align}
under the assumption that $\Braket{\Psi|\hat{Q}^2|\Psi}$ is finite. 
Let us confirm that the operators $\left(\hat{Q},\hat{P}\right)$ satisfy the requirements. The commutation relationship can be checked as follows:
\begin{align}
 \left[\hat{Q},\hat{P}\right]&=\frac{1}{\Braket{\Psi|\hat{Q}^2|\Psi}}\int d^d\bm{ x}d^d\bm{y}d^d\bm{z}d^d\bm{w}\, v(\bm{x})\left[\hat{\Gamma}(\bm{x}),\hat{\Gamma}(\bm{y})^{\mathrm{T}}\right]\left(-\Omega(\bm{y},\bm{z})\right)M\left(\bm{z},\bm{w}\right)v(\bm{w})\nonumber\\
 &= i \frac{1}{\Braket{\Psi|\hat{Q}^2|\Psi}}\int d^d\bm{x}d^d\bm{y}v (\bm{x})M(\bm{x},\bm{y})v(\bm{y})=i.
\end{align}
The off-diagonal elements of covariance matrix vanish:
\begin{align}
& \mathrm{Re}\left(\Braket{\Psi|\hat{Q}\hat{P}|\Psi}\right)\nonumber\\
&=\frac{1}{\Braket{\Psi|\hat{Q}^2|\Psi}}\int d^d\bm{x}d^d\bm{y}\bm{z}d^d\bm{w}\, v(\bm{x})^{\mathrm{T}} \mathrm{Re}\left(\Braket{\Psi|\hat{\Gamma}(\bm{x})\hat{\Gamma}(\bm{y})^{\mathrm{T}}|\Psi}\right)\left(-\Omega(\bm{y},\bm{z})\right)M(\bm{z},\bm{w})v(\bm{w})\nonumber\\
 &=\frac{1}{\Braket{\Psi|\hat{Q}^2|\Psi}}\int d^d\bm{x}d^d\bm{y}d^d\bm{z}d^d\bm{w}\, v(\bm{x})^{\mathrm{T}} M(\bm{x},\bm{y})\left(-\Omega(\bm{y},\bm{z})\right)M(\bm{z},\bm{w})v(\bm{w})=0, 
\end{align}
where we have used the fact that $\Omega(\bm{x},\bm{y})$ is anti-symmetric in the last line. 
The last one is the expectation value of $\hat{P}$:
\begin{align}
& \Braket{\Psi|\hat{P}^2|\Psi}\nonumber\\
&=\frac{1}{\Braket{\Psi|\hat{Q}^2|\Psi}^2}\int d^d\bm{x}_1\cdots d^d\bm{x}_6\, v(\bm{x}_1)^{\mathrm{T}}M(\bm{x}_1,\bm{x}_2)\Omega(\bm{x}_2,\bm{x}_3)\mathrm{Re}\left(\Braket{\Psi|\hat{\Gamma}(\bm{x}_3)\hat{\Gamma}(\bm{x}_4)^{\sps{T}}|\Psi}\right)\nonumber\nonumber\\
&\quad \quad\quad \quad \quad\quad\quad \quad\quad\times\left(-\Omega(\bm{x}_4,\bm{x}_5)\right)M(\bm{x}_5,\bm{x}_6)v(\bm{x}_6)\nonumber\\
&=\frac{1}{\Braket{\Psi|\hat{Q}^2|\Psi}^2}\int d^d\bm{x}_1d^d\bm{x}_4d^d\bm{x}_5 d^d\bm{x}_6\, v(\bm{x}_1)^{\mathrm{T}}\frac{1}{4}\Omega(\bm{x}_1,\bm{x}_4)\left(-\Omega(\bm{x}_4,\bm{x}_5)\right)M(\bm{x}_5,\bm{x}_6)v(\bm{x}_6)\nonumber\\
 &=\frac{1}{4}\frac{1}{\Braket{\Psi|\hat{Q}^2|\Psi}}, 
\end{align}
where we have assumed that
\begin{align}
 \int d^d\bm{x}_2 d^d\bm{x}_3\,  M(\bm{x}_1,\bm{x}_2)\Omega(\bm{x}_2,\bm{x}_3)M(\bm{x}_3,\bm{x}_4)=\frac{1}{4}\Omega\left(\bm{x}_1,\bm{x}_4\right)\label{eq_mOmegam_con}
\end{align}
holds. Therefore, $\det{m}=\frac{1}{4}$ is satisfied.
Eq.~(\ref{eq_mOmegam_con}) is the continuum limit of Eq.~(\ref{eq_mOmegam_dis}). If the Gaussian state $\ket{\Psi}$ can be obtained from a Gaussian state in a HO chain in an appropriate continuum limit, this equation holds. For example, it is possible to confirm that the vacuum state of the free scalar field theory is a Gaussian state satisfying Eq.~(\ref{eq_mOmegam_con}).

\section{Example: Scrambling of information in discretized scalar field theory}\label{sec_ex}
In this section, we calculate the time-evolution of QIC operators in descretized scalar field theory in a flat $(1+1)$-dimensional spacetime. Imposing a periodic boundary condition on the scalar field, the free Hamiltonian is given by
\begin{align}
 \hat{H}=\frac{1}{2}\int_{-L/2}^{L/2}dx \, :\hat{\Pi}(x)^2:+\frac{1}{2}\int_{-L/2}^{L/2}:\left(\partial_x\hat{\phi}(x)\right)^2:+\frac{m^2}{2}\int_{-L/2}^{L/2}:\hat{\phi}(x)^2:.
\end{align}
Here, $L$ denotes the entire length of the space, $m$ is the mass of the field, and $:\hat{\mathcal{O}}:$ is the normal ordering of a linear operator $\hat{\mathcal{O}}$. The canonical operators $\{(\hat{q}_n,\hat{p}_n)\}_{n=1}^N$ for the descretized theory corresponds to
\begin{align}
 \hat{\phi}(x)\to \frac{1}{\sqrt{m\epsilon}}\hat{q}_n, \quad \hat{\Pi}(x)\to \sqrt{\frac{m}{\epsilon}}\hat{p}_n,
\end{align}
where $\epsilon \equiv L/N$ is the lattice spacing. The descritized Hamiltonian is given by
\begin{align}
 \hat{H}=\frac{1}{2}\sum_{n=1}^N:\hat{p}_n^2:+\left(\frac{1}{2}+\eta\right)\sum_{n=1}^N:\hat{q}_n^2:-\eta \sum_{n=1}^N\hat{q}_n:\hat{q}_{n+1}:,
\end{align}
where $\eta\equiv 1/(m\epsilon)^2$. The Hamiltonian can be re-written as
\begin{align}
 \hat{H}=\sum_{k=1}^N\omega_k\hat{a}_k^\dag\hat{a}_k,
\end{align}
where 
\begin{align}
 \omega_k\equiv \sqrt{1+2\eta\left(1-\cos\left(\frac{2\pi k}{N}\right)\right)}
\end{align}
and the creation and annihilation operators satisfy
\begin{align}
 \hat{q}_n=\sum_{k=1}^N\frac{1}{\sqrt{2\omega_k}}\left(\hat{a}_kf_k(n)+\hat{a}_k^\dag f_k(n)^*\right),\quad \hat{p}_n=\frac{1}{i}\sum_{k=1}^N \sqrt{\frac{\omega_k}{2}}\left(\hat{a}_kf_k(n)-\hat{a}_k^\dag f_k(n)^*\right)\label{eq_ra}
\end{align}
with the mode functions
\begin{align}
 f_k(n)\equiv \frac{1}{\sqrt{N}}\exp{\left(2\pi i k \frac{n}{N}\right)}.
\end{align}
Introducing a $2N\times 2N$ matrix $A$ whose elements are given by
\begin{align}
 A_{ab}\equiv
\begin{cases}
 \frac{1}{\sqrt{2\omega_{(b+1)/2}}}f_{(b+1)/2}(\frac{a+1}{2})&\quad \left(a,b:\text{odd}\right)\\
\frac{1}{\sqrt{2\omega_{b/2}}}f_{b/2}^*(\frac{a+1}{2})&\quad \left(a:\text{odd},\, b:\text{even}\right)\\
\frac{1}{i}\sqrt{\frac{\omega_{(b+1)/2}}{2}}f_{(b+1)/2}(\frac{a}{2})& \quad \left(a:\text{even},\,b :\text{odd}\right)\\
-\frac{1}{i}\sqrt{\frac{\omega_{b/2}}{2}}f_{b/2}^*(\frac{a}{2})& \quad \left(a,b:\text{even}\right)
\end{cases},
\end{align}
Eq.~(\ref{eq_ra}) is written as $\hat{\bm{r}}=A\left(\hat{a}_1,\hat{a}_1^\dag,\cdots,\hat{a}_N,\hat{a}_N^\dag\right)^\mathrm{T}$.

Suppose that we imprint information of $\theta $ by using a shift operator $\hat{W}(\theta)$ defined Eq.~(\ref{eq_qlinear}) at $t=0$. Assuming that the system is in a Gaussian state $\ket{\Psi}$ before the write operation, the wighting functions of conjugate QIC operator are given by Eq.~(\ref{eq_u}). At $t>0$, the system evolves into $\hat{U}(t)\hat{W}(\theta)\ket{\Psi}$, where $\hat{U}(t)\equiv e^{-i\hat{H}t}$ is the unitary evolution operator. As is presented in \cite{YWH}, the QIC operators at $t>0$ are given by
\begin{align}
 \hat{Q}(t)\equiv \hat{U}(t)\hat{Q}\hat{U}(t)^\dag,\quad \hat{P}(t)\equiv \hat{U}(t)\hat{P}\hat{U}(t)^\dag.
\end{align}
In terms of weighting functions, they are written as $\hat{Q}(t)=\bm{v}(t)^{\mathrm{T}}\hat{\bm{r}}$ and $\hat{P}(t)=\bm{u}(t)^{\mathrm{T}}\hat{\bm{r}}$, where we have defined
\begin{align}
 \bm{v}(t)^\mathrm{T}&\equiv \bm{v}^{\mathrm{T}}A\bigoplus_{k=1}^N
\begin{pmatrix}
 e^{i\omega_k}&0\\
 0&e^{-i\omega_k} 
\end{pmatrix} 
A^{-1},\quad \bm{u}(t)^\mathrm{T}\equiv \bm{u}^{\mathrm{T}}A\bigoplus_{k=1}^N
\begin{pmatrix}
 e^{i\omega_k}&0\\
 0&e^{-i\omega_k} 
\end{pmatrix} 
A^{-1}.
\end{align}

\begin{figure}[htbp]
\includegraphics[width=7.5cm]{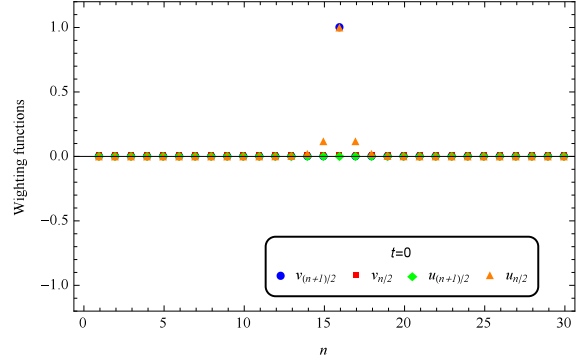}
\caption{The wighting functions of QIC operators at $t=0$.}
\label{fig_g0}
\end{figure}

\begin{figure}[htbp]
\includegraphics[width=7.5cm]{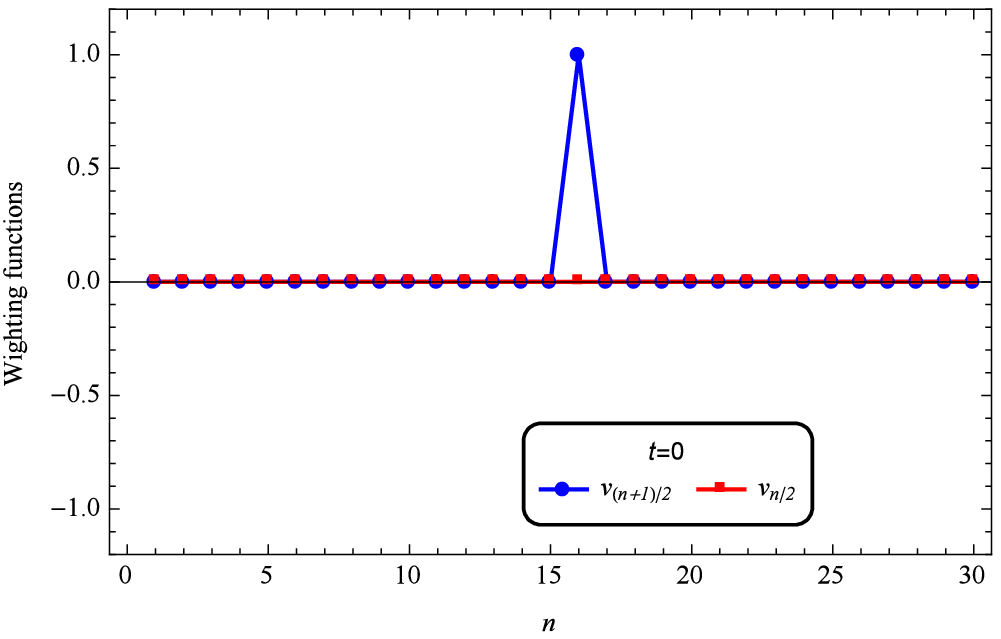}
\caption{The wighting functions of $\hat{Q}(0)$ with lines.}
\label{fig_g0q}
\end{figure}

\begin{figure}[htbp]
\includegraphics[width=7.5cm]{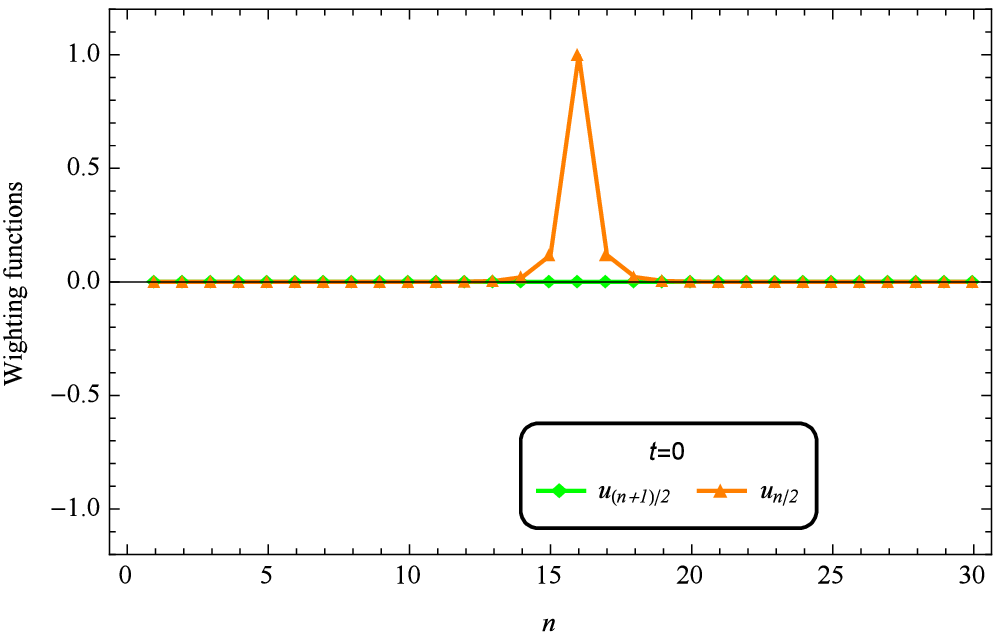}
\caption{The wighting functions of $\hat{P}(0)$ with lines.}
\label{fig_g0p}
\end{figure}

\begin{figure}[htbp]
\includegraphics[width=7.5cm]{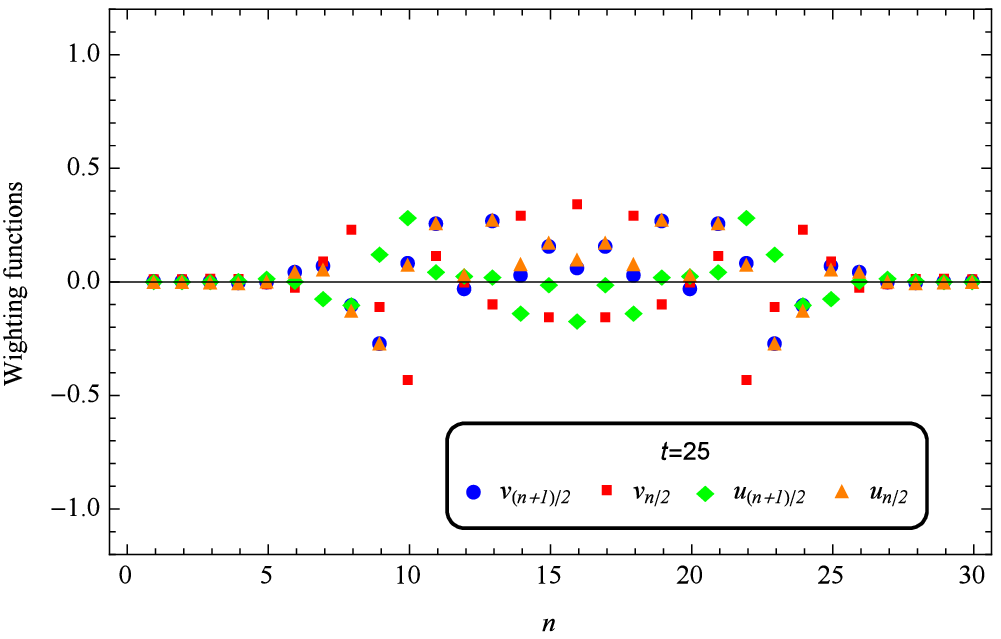}
\caption{The wighting functions of QIC operators at $t=25$.}
\label{fig_g25}
\end{figure}

\begin{figure}[htbp]
\includegraphics[width=7.5cm]{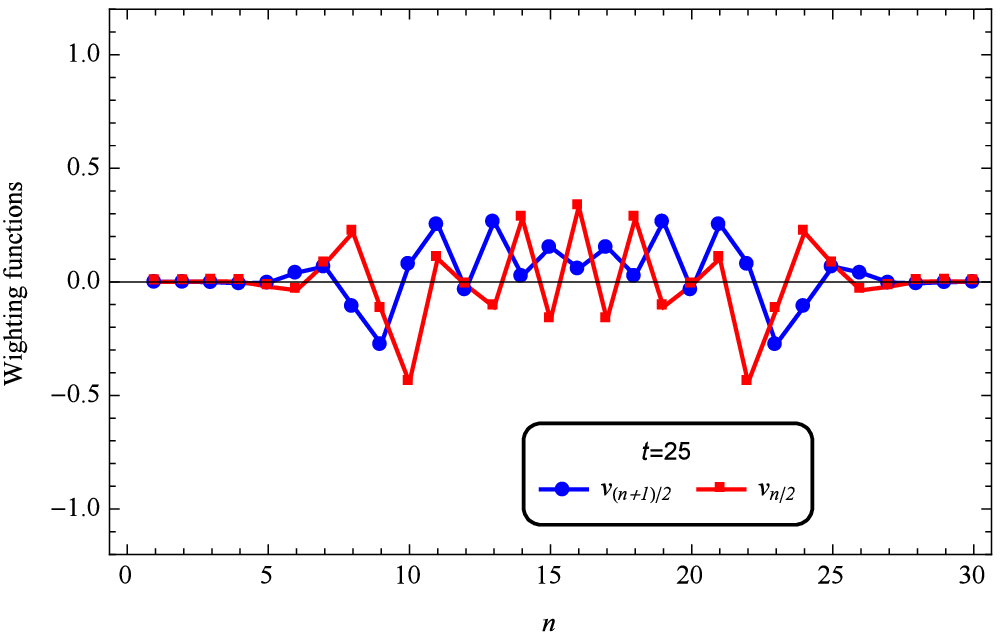}
\caption{The wighting functions of $\hat{Q}(25)$ with lines.}
\label{fig_g25q}
\end{figure}

\begin{figure}[htbp]
\includegraphics[width=7.5cm]{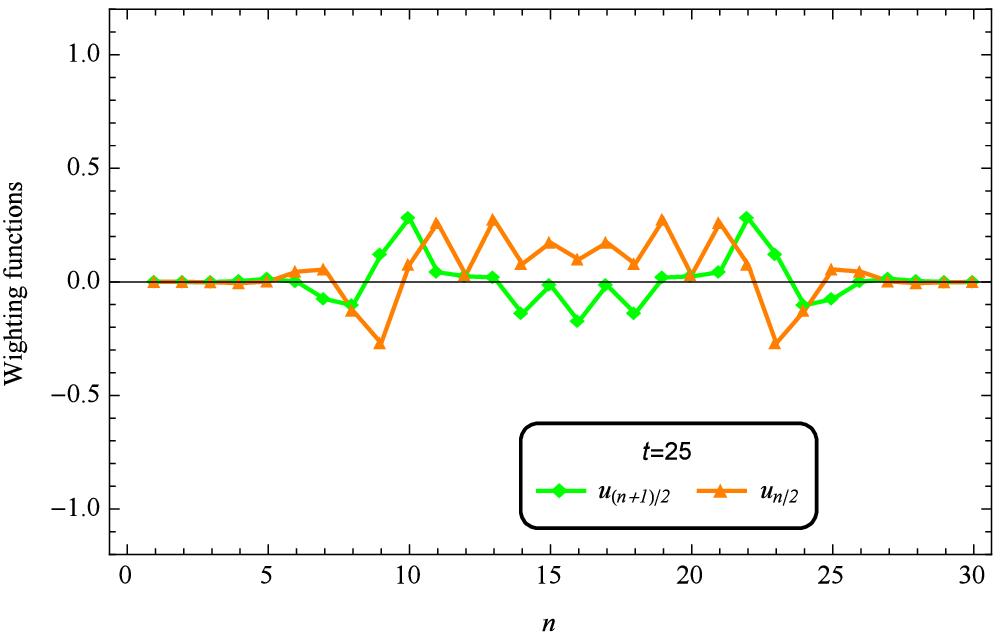}
\caption{The wighting functions of $\hat{P}(25)$ with lines.}
\label{fig_g25p}
\end{figure}

\begin{figure}[htbp]
\includegraphics[width=7.5cm]{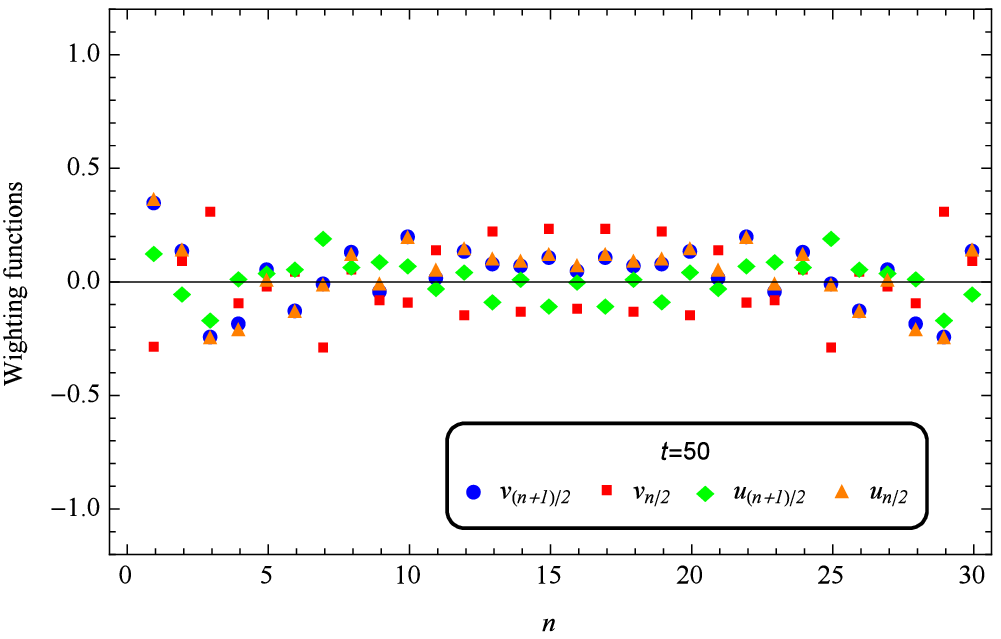}
\caption{The wighting functions of QIC operators at $t=50$.}
\label{fig_g50}
\end{figure}

\begin{figure}[htbp]
\includegraphics[width=7.5cm]{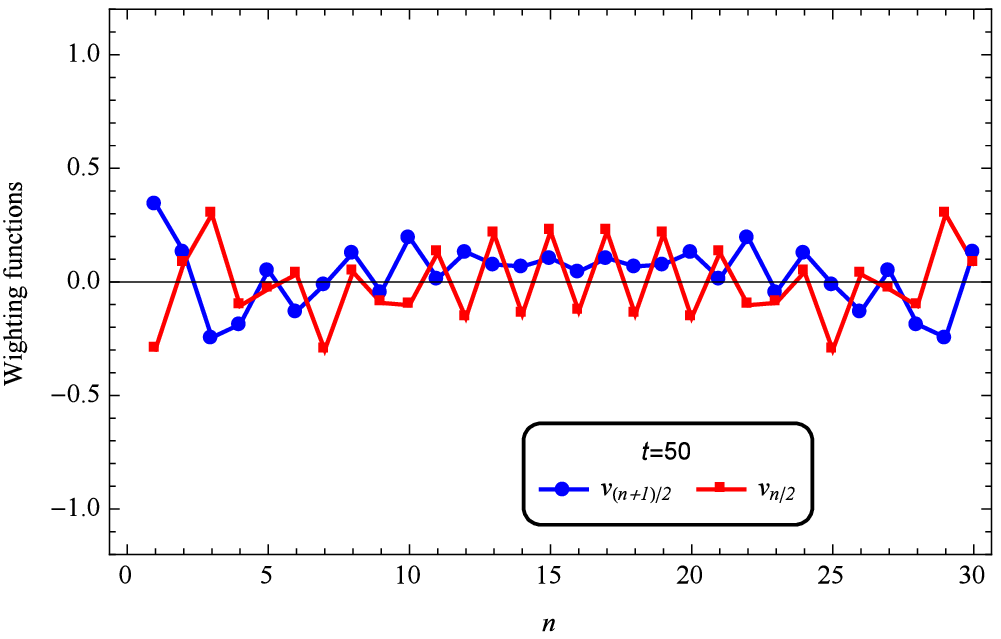}
\caption{The wighting functions of $\hat{Q}(50)$ with lines.}
\label{fig_g50q}
\end{figure}

\begin{figure}[htbp]
\includegraphics[width=7.5cm]{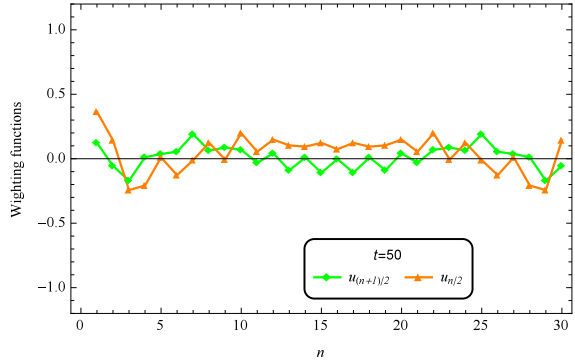}
\caption{The wighting functions of $\hat{P}(50)$ with lines.}
\label{fig_g50p}
\end{figure}

As an example, the evolution of weighting functions for the vacuum state is shown in Figs.~(\ref{fig_g0})-(\ref{fig_g50p}). The parameters are fixed as $N=30$ and $\eta=0.4$, and the shift operator is characterized by $\hat{Q}=\hat{q}_{15}$. 
Even at $t=0$, the weighting functions for $\hat{P}$ take non-zero values even for $n\neq 15$ due to the entanglement-induced information delocalization, although they are localized around $n=15$. At $t=25$, due to the free time-evolution of the system, the information starts to propagate through the system. The weighting functions are almost zero at the sites far from $n=15$, which is consistent with the intuition that the there is a limitation for the speed of information propagation. At $t=50$, the profiles of weighting functions show that the QIC mode is spreaded over the whole system.
The complexity of the QIC operators seems to increase in time, which may be related to measures of scrambling and quantum chaos such as out-of-order correlators \cite{MS} and  tripartite information \cite{TPI}. A quantative analysis on the scrambling effect by using the evolution of QIC operators is beyond the scope of this paper and left for future research. 


\section{Summary}\label{sec_summary}
In this paper, we extended the pictures of partners and a QIC storing the information imprinted into a systems. The results presented in Section \ref{sec_fin_dim} show that for a finite-dimensional system, a QIC always exists for general write operations. When there is no criterion that singles out a QIC, non-equivalent QIC can be constructed, implying that there are many ways to process the imprinted information. In Section \ref{sec_inf_dim}, we extended the proof of existence of a QIC to CV systems. For shift write operations, a unique QIC always exists for an arbitrary Gaussian state, imposing the condition that the QIC mode is characterized by linear combinations of canonical variables. 

In Section \ref{sec_ex}, the time-evolution of QIC operators is calculated for the discretized scalar field theory. The wighting functions of QIC operators are spread over the system due to the free evolution. Fig~(\ref{fig_g50}) suggests that we need to prepare a highly lon-local detector in order to detect the QIC mode at the late time. It would be interesting to investigate the scrambling effect by using QICs in future research.

\appendix
\section{The SWAP operation for qudits}\label{app_swap}
The SWAP operation for a two-qudit system is given by
\begin{align}
 \hat{U}_{\mathrm{swap}}=\sum_{i,j=1}^d\ket{i}\bra{j}\otimes \ket{j}\bra{i},
\end{align}
where $\left\{\ket{i}\right\}_{i=1}^d$ is an orthonormal basis for a single-qudit Hilbert space $\mathcal{H}_d$. This operator swaps the state of first qudit for that of the second qudit. That is, for arbitrary unit vector $\ket{\phi}\ket{\psi}\in\mathcal{H}_d$, it holds
\begin{align}
 \hat{U}_{\mathrm{swap}}\ket{\phi}\ket{\psi}=\ket{\psi}\ket{\phi}.
\end{align}
Since $\hat{U}_{\mathrm{swap}}$ is a linear operator, it can be expressed as
\begin{align}
 \hat{U}_{\mathrm{swap}}=\sum_{\mu,\nu=0}^{d^2-1}c_{\mu\nu}\hat{t}_{\mu}\otimes \hat{t}_{\nu},
\end{align}
where $c_{\mu\nu}\in\mathbb{C}$, $\hat{t}_0\equiv \mathbb{I}_d$ and $\left\{\hat{t}_{i}\right\}_{i=1}^{d^2-1}$ is a basis of $\mathfrak{su}(d)$ algebra satisfying $\mathrm{Tr}_{\mathcal{H}_d}\left(\hat{t}_{i}\hat{t}_{j}\right)=d\delta_{ij}$. Since $\mathrm{Tr}_{\mathcal{H}_d}\left(\hat{t}_{\mu}\hat{t}_{\nu}\right)=d\delta_{\mu\nu}$ holds for $\mu,\nu=0,\cdots, d^2-1$, we get
\begin{align}
 c_{\mu\nu}&=\frac{1}{d^2}\mathrm{Tr}_{\mathcal{H}_d^{\otimes 2}}\left(\hat{U}_{\mathrm{swap}}\hat{t}_{\mu}\otimes \hat{t}_{\nu}\right)\nonumber\\
 &=\frac{1}{d^2}\sum_{i,j=1}^d\Braket{j|\hat{t}_{\mu}|i}\Braket{i|\hat{t}_{\nu}|j}\nonumber\\
 &=\frac{1}{d^2}\mathrm{Tr}_{\mathcal{H}_d}\left(\hat{t}_{\mu}\hat{t}_{\nu}\right) =\frac{1}{d}\delta_{\mu\nu}.
\end{align}
Therefore, the SWAP operation is expressed in terms of a basis of $\mathfrak{su}(d)$ as
\begin{align}
 \hat{U}_{\mathrm{swap}}=\frac{1}{d}\sum_{\mu=0}^{d^2-1}\hat{t}_{\mu}\otimes\hat{t}_{\mu}.
\end{align}

\begin{acknowledgments}
The authors thank Ursula Carow-Watamura, Takeshi Tomitsuka, Naoki Watamura, and Satoshi Watamura for useful discussions. This research was partially supported by JSPS KAKENHI Grant Numbers JP16K05311 (M.H.) and JP18J20057 (K.Y.), and by Graduate Program on Physics for the Universe of Tohoku University (K.Y.).
\end{acknowledgments}


\begin{thebibliography}{99}
\bibitem{HR} S. W. Hawking, \emph{Particle creation by black holes}, Commun. Math. Phys. \textbf{43}, 199 (1975).

\bibitem{BHIP}S. W. Hawking, \emph{Breakdown of predictability in gravitational collapse}, Phys. Rev. D \textbf{14}, 2460 (1976).

\bibitem{HSU}M. Hotta, R. Sch\"utzhold, and W. G. Unruh, \emph{Partner particles for moving mirror radiation and black hole evaporation}, Phys. Rev. D \textbf{91}, 124060 (2015).

\bibitem{TYH}J. Trevison, K. Yamaguchi, M. Hotta, \emph{Spatially Overlapped Partners in Quantum Field Theory}, J. Phys. A: Math. Theor. in press (2019).

\bibitem{YWH} K. Yamaguchi, N. Watamura, and M. Hotta, \emph{Quantum Information Capsule and Information Delocalization by Entanglement in Multiple-qubit Systems}, Phys. Lett. A. \textbf{383}, 1255 (2019).

\bibitem{CS1} D. Gross and J. Eisert, \emph{Novel Schemes for Measurement-Based Quantum Computation}, Phys. Rev. Lett. \textbf{98}, 220503 (2007).

\bibitem{CS2} J.-M. Cai, W. Dx\"ur, M. Van den Nest, A. Miyake, and H. J. Briegel, \emph{Quantum Computation in Correlation Space and Extremal Entanglement}, Phys. Rev. Lett. \textbf{103}, 050503 (2009).

\bibitem{E1} M. Gring, M. Kuhnert, T. Langen, T. Kitagawa, B. Rauer, M. Schreitl, I. Mazets, D. Adu Smith, E. Demler, J. Schmiedmayer, \emph{Relaxation and Prethermalization in an Isolated Quantum System}, Science \textbf{337}, 1318 (2012).

\bibitem{E2} T. Langen, S. Erne, R. Geiger, B. Rauer, T. Schweigler, M. Kuhnert, W. Rohringer, I. E. Mazets, T. Gasenzer, and J. Schmiedmayer, \emph{Experimental observation of a generalized Gibbs ensemble}, Science \textbf{348}, 207 (2015).

\bibitem{E3} R. Islam, R. Ma, P. M. Preiss, M. E. Tai, A. Lukin, M. Rispoli, and M. Greiner, \emph{Measuring entanglement entropy in a quantum many-body system}, Nature \textbf{528}, 77, (2015).

\bibitem{E4} A. M. Kaufman, M. E. Tai, A. Lukin, M. Rispoli, R. Schittko, P. M. Preiss, and M. Greiner, \emph{Quantum thermalization through entanglement in an isolated many-body system}, Science \textbf{353}, 794 (2016) .

\bibitem{QF}C. W. Helstrom, \emph{Minimum mean-squared error of estimates in quantum statistics}, Phys. Lett. A \textbf{25}, 101 (1967).

\bibitem{NC} M. A. Nielsen and I. L. Chuang, \emph{Quantum Computation and Quantum Information}, Cambridge University Press (2000).

\bibitem{BP} S. L. Braunstein and A. K. Pati, \emph{Quantum Information Cannot Be Completely Hidden in Correlations: Implications for the Black-Hole Information Paradox}, Phys. Rev. Lett. \textbf{98}, 080502 (2007).

\bibitem{MPSS} K. Modi, A. K. Pati, A. Sen(De), and U. Sen, \emph{Masking Quantum Information is Impossible}, Phys. Rev. Lett. \textbf{120}, 230501 (2018).

\bibitem{U}W. G. Unruh, \emph{Notes on black-hole evaporation}, Phys. Rev. D \textbf{14}, 870 (1976).

\bibitem{dW}  B. De Witt, \emph{Quantum gravity: the new synthesis}, General Relativity: An Einstein CentenarySurvey, edited by S.W. Hawking and W. Israel (Cambridge University Press, Cambridge,England), 680 (1979).

\bibitem{EE} A. S. Holevo, M. Sohma, and O. Hirota, \emph{Capacity of quantum Gaussian channels}, Phys. Rev.A \textbf{59}, 1820 (1999).

\bibitem{QCV} See e.g., A. Serafini, \emph{Quantum Continuous Variables: A Primer of Theoretical Methods}, CRCPress, 2017.

\bibitem{MS} J. Maldacena, S. H. Shenker, and D. Stanford, \emph{A bound on chaos}, J. High Energ. Phys. \textbf{1608}, 106 (2016).

\bibitem{TPI} P. Hosur, X.-L. Qi, D. A. Roberts, B. Yoshida, \emph{Chaos in quantum channels}, J. High Energ. Phys. \textbf{1602}, 004 (2016).

\end{thebibliography}
\end{document}